\documentclass[12pt]{article}
		\title{A Bayesian Search for the Higgs Particle}

\pdfoutput=1
\date{}
	\usepackage{amsmath,latexsym, graphics,graphicx,showexpl,amsthm,amsfonts,fullpage}
\usepackage{graphicx}
\usepackage{caption}
\usepackage{subcaption}
\usepackage{bbm} 
\usepackage{natbib}
\usepackage[titletoc]{appendix}
\usepackage{epstopdf}
\usepackage[pdftex,bookmarks,colorlinks=true,linkcolor=black]{hyperref}
\usepackage{algpseudocode,algorithm}
\usepackage{authblk}
\usepackage{multicol}
\usepackage{hyperref}
\makeatother
\author[1]{Shirin Golchi\thanks{e-mail: golchi.shirin@gmail.com}}
\author[2]{Richard Lockhart}

\affil[1]{University of British Columbia - Okanagan, Statistics}
\affil[2]{Simon Fraser university, Department of Statistics and Actuarial Science}

\begin{document}
	\maketitle

\linespread{1}
\begin{abstract}
The statistical procedure used in the search for the Higgs boson is investigated in this paper. A Bayesian hierarchical model is proposed that uses the information provided by the theory in the analysis of the data generated by the particle detectors. In addition, we develop a Bayesian decision making procedure that combines the two steps of the current method (discovery and exclusion) into one and can be calibrated to satisfy frequency theory error rate requirements. 
 
\end{abstract}

\noindent%

\section{Introduction}
\label{sec:intro}

The Standard Model (SM) of particle physics is a theory that describes the dynamics of subatomic particles. The Higgs particle is an essential component of the SM; its existence explains why other elementary particles are massive \citep{ 1964PhRvL..13..321E, 1964PhRvL..13..508H, PhysRevLett.13.585, 2009IJMPA..24.2601G}. The existence of the Higgs boson needs to be confirmed by experiment. The Large Hadron Collider (LHC) at the European organization for nuclear research, known as CERN, is a high energy collider specifically designed and constructed to detect the Higgs particle. 
Two beams of protons circulating at very high speeds in the LHC collide inside two detectors (ATLAS and CMS). Collisions between a proton in one beam and a proton in the other beam result in generation of new particles, possibly including the Higgs boson; each such collision is an event. Some of the particles generated can be tracked and measured in the detectors. 
However, the Higgs particle, if generated, decays extremely quickly into other known SM particles and cannot be detected directly.  Instead, the existence of the Higgs particle must be inferred by looking for those combinations of  detectable particles that are predicted by the SM.

Once a Higgs particle has been created (by one of several `production mechanisms') in a proton-proton collision,
there are several different processes, called `decay modes', through which the particle may decay. The decay process can be reconstructed based on the detected collision byproducts. Events with reconstructed processes that match one of the possible Higgs decay modes and pass other selection criteria (called cuts) are recorded as ``Higgs candidates" and the invariant mass of the unobserved particle is computed from the reconstruction.  A histogram of the estimator of the mass is then created for each decay mode (some analyses including $H\rightarrow 2$ photons use unbinned likelihood fits instead of histograms).  However, there are other processes, not involving the Higgs boson, that can result in the generation of Higgs event byproducts which also pass the cuts; these are called background events. Thus the histogram created is either a mixture of background events and events in which a Higgs particle was created or just a histogram of background events if the Higgs particle does not exist.

Luckily, the SM predicts, as a function of the mass of the Higgs particle and the energy of the colliding beams, the expected rate at which  events generate Higgs particles (a quantity proportional to the so-called Higgs cross section). It also predicts, as a function of the same parameters, the probability that the particle will decay by a given decay mode and produce byproducts which pass the cuts.  Effectively then, the SM predicts that if the Higgs particle exists and has mass $m_H$ then there will be a bump on the histogram of invariant masses whose size and shape are completely predicted by unknown mass, $m_H$, and other measured quantities such as beam energy.  The statistical problem is to determine whether or not such a bump exists and if so at what mass it is centred.

The search procedure has two stages. The first stage is discovery, or search while the 
second is called exclusion.  
For each mass in a range of possible masses the null hypothesis that there is no Higgs particle is tested against the alternative that it exists and has this specific mass.  A p-value, called a local p-value, is computed for each mass; 
if the smallest p-value is less than $1-\Phi(5)$ (a 5$\sigma$ effect---$\Phi$ is the standard normal cumulative density function) then the particle is declared to have been `observed'. If not then the second stage, exclusion, is pursued.  For each mass in the range under consideration the null hypothesis that
the particle exists, at this mass, with the predicted cross section, is tested against a background only null hypothesis.

While the main goal is to discover the Higgs particle and the parameter of interest is the mass of the Higgs boson, denoted by $m_H$, the local hypothesis tests (which are likelihood ratio tests) focus on the cross section at each mass.  The background model parameters and any other unknown parameters of the signal function are treated as nuisance parameters. 

 In the searches for the Higgs particle described in \cite{phys1} and \cite{phys2}, in the discovery phase, theoretical predictions are used to define the signal function for different production mechanisms and analysis categories. An overall cross section parameter in form of a unitless scaling factor, generally denoted by $\mu$, is used in the model.  The standard model predicts that $\mu=1$ and the null hypothesis that there is no Higgs is represented by $\mu=0$. The procedures described in \cite{phys1} and \cite{phys2} treat the alternative to $\mu=0$ as $\mu>0$. Treating $\mu$ as a free parameter potentially increases the possibility of observing signal at more than one mass value. This is noted in reporting 2-d confidence intervals that are made for the mass of the Higgs particle and the cross section parameter where a range of pairs of mass values and cross sections are considered consistent with the signal \cite{Phys3}.

In order to fix some notation we now give some further details of the two stages for the current practice. Some key weaknesses of the existing method are highlighted in the process.

{\bf Discovery:} For each mass value, $m\in(m_0,m_n)$, in the search window, a local likelihood ratio test (see \cite{Cowan11} for details) is performed at the significance level, $\alpha_0=1-\Phi(5)=3\times 10^{-7}$,
\begin{equation}
H_0: \mu=0\hskip 20pt \text{vs.} \hskip 20pt H_A: \mu>0.
\end{equation}
Note that the cross section predicted by the SM is not used under the alternative. Discovery is announced if the p-value associated with any of the local tests (the minimum local p-value) is below $\alpha_0$ and the mass of the Higgs particle is estimated as the corresponding mass value. We note that this is not the final estimate of the mass; a further detailed analysis of the decay products in the relevant events is used to provide an estimate with uncertainties.  The plot of local p-values does not itself provide an interval estimate for the mass if the particle is detected.

This testing procedure is equivalent to rejecting the null hypothesis if any of  a family of local test statistics, $\nu(m)$, (e.g., log likelihood ratios) indexed by the unknown mass $m$ is larger than a predefined level $\kappa$.  \cite{GrossVitells10} proposed a method for estimating the ``global p-value" of this testing procedure as the null probability that the maximum of the local test statistics is greater than the observed maximum. Their method is based on the results of \cite{Davies87} and gives:
\begin{align}
P_G&=P_{H_0}(\max_m \nu(m)>\kappa)\nonumber\\
&\approx P_{H_0}(\chi^2_d>\kappa)+\text{E}_{H_0} \left(N(\kappa)\right)
\end{align}
where $\kappa$ in this case is the observed maximum statistic, $\chi^2_d$, denoting a chi-squared distribution with $d$ degrees of freedom, is the null distribution of $\nu(m)$ and $\text{E}_{H_0} \left(N(\kappa)\right)$ is the expected number of upcrossings of the level $\kappa$ by the process $\nu(m)$. Using the method of \cite{GrossVitells10}  one can estimate the (global) type I error rate associated with the discovery procedure that is based on the local tests,
\begin{align}
\alpha_G&=P_{H_0}(\max_m \nu(m)>\kappa_{\alpha_0})\nonumber\\
&\approx P_{H_0}(\chi^2_d>\kappa_{\alpha_0})+\text{E}_{H_0} \left(N(\kappa_{\alpha_0})\right)\nonumber\\
&=\alpha_0+\text{E}_{H_0} \left(N(\kappa_{\alpha_0})\right),
\end{align}
where $\kappa_{\alpha_0}$ is the $\chi^2_d$ quantile corresponding to $\alpha_0$. Since $\text{E}_{H_0} \left(N(\kappa)\right)\geq 0$, the actual global type I error rate is larger than the controlled local type I error rates.  The size of the difference depends on the specific statistical model.

{\bf Exclusion:}
In the second stage, carried out only if no particle is detected at the first stage, further investigation is done to exclude regions of $m$ which are unlikely values for the mass of the Higgs boson. The theoretical cross section is tested at  significance level $\alpha_2=0.05$ at the exclusion step \citep{Cowan11},
\begin{equation}
H_0: \mu=1\hskip 20pt \text{vs.} \hskip 20ptH_A: \mu<1.
\end{equation}
Any mass whose corresponding theoretical cross section is rejected is excluded from the range of possible masses for the Higgs particle.

A number of questions/criticisms arise regarding the existing procedure:
Why is the cross section parameter treated as unknown and estimated in the discovery stage while the theoretical value of this parameter is used in the exclusion? In other words, why is a smaller cross section than the one predicted by theory eligible at the discovery step when it might well be excluded when testing for exclusion? The global type I error rate associated with the local discovery procedure is larger than the local thresholds and is not obtained. What would the decision be if more than one local p-value for well separated mass values fall below the local significance level at the discovery step? This scenario is more likely when the cross section is treated as unknown.  Current procedure sets a very low Type I error rate for each test and yields a procedure with an unknown global type I error rate.  The suggestion that this error rate be computed only for data sets crossing some local error rate threshold guarantees that the estimated local error rate has no straightforward frequency theory interpretation.

In the following we propose a Bayesian procedure that combines discovery and exclusion into a single decision making step and enables the calculation and control of various frequency theory global error rates. This is done by specifying some elements of the decision making procedure (priors or loss values) and calibrating the rest of the elements to satisfy predetermined error rates. Our procedure takes advantage of the information provided by theory but can be modified to treat the cross section as an unknown parameter. In other words the proposed procedure is capable of consistently using the information provided by the theory as well as incorporating any uncertainty that might be associated with this information. Unlike the current practice, which could potentially result in different decisions in the discovery and exclusion steps, our procedure provides a set including all the possible mass values for the Higgs particle. 

The rest of the paper is organized as follows. In Section 2, we introduce a Bayesian hierarchical model and describe inference for model parameters. Section 3 is dedicated to the Bayesian decision making procedure that, while combining the discovery and exclusion steps, can be calibrated to obtain desired frequency-theory error-rates. In Section 4, the proposed inference and decision making methods are applied to simulated Higgs data provided to us by the CMS Higgs group. Section 5 follows with concluding remarks.

\section{Model and inference}
\label{model}
In this section we introduce a Bayesian hierarchical model that captures the features of the Higgs problem. A sequential Monte Carlo (SMC) algorithm is used to make inference about the model parameters.

Suppose that the data, i.e., the invariant masses recorded by the detector, are realizations of a Poisson process whose intensity function is given by the sum of a background process $\Lambda(m)$ and a signal function $s_{m_H}(m)$. The shape of the signal function is known and its location is determined by the unknown parameter, $m_H\in {\cal M}$, where ${\cal M}=\{\emptyset\}\cup (m_0,m_n)$ ($(m_0,m_n) \subset \mathcal{R}^+-\{0\}$, i.e., $m_0$ and $m_n$ are strictly positive). The parameter, $m_H$, is the unknown mass of the Higgs particle where $m_H\in (m_0,m_n)$ means that the Higgs boson has a mass in the search window, $(m_0,m_n)$, while $m_H=\emptyset$ refers to the case that  the particle does not exist, at least not with a mass in $(m_0,m_n)$. We model the uncertainty about the background $\Lambda(m)$ as a log-Gaussian process,
\begin{equation}
\label{backprior}
\log\Lambda_{\eta,\sigma^2}(m)\sim {\cal GP}(\xi(m),\rho_{\eta,\sigma_2}(m,m')), \hskip 10pt m\in (m_0,m_n).
\end{equation}
with a known mean function, $\xi(m)$, and covariance function given by,
\begin{equation}
\label{covfnct}
\rho_{\eta,\sigma_2}(m,m')=\sigma^2\exp(-\eta(m-m')^2),
\end{equation}
where $\sigma^2$ is the variance parameter and $\eta$ is the correlation parameter that controls the smoothness of the background function. The notation $\Lambda_{\eta,\sigma^2}$ is used to show the dependence of the background function on the covariance parameters. For the sake of brevity we drop the subscript from here on. 

The functional form of the background model and the uncertainties associated with it are typically obtained from Monte Carlo studies \citep{Phys4}. The mean function in the background prior in (\ref{backprior}) would ideally  be determined in the same way. However, the log-Gaussian process is a realistic model that allows the background function to deviate from the mean function while letting the data correct for a possibly incorrect prior mean.

We choose the signal function as a Gaussian probability density function with the location parameter $m_H$ (in the current practice \citep{phys1, phys2} a slightly more complex signal shape called the ``crystal ball function" is used). Thus, the signal function is given by
\begin{equation}
\label{signal}
s_{m_H}(m)=c_{m_H}\hskip 2pt\phi(\frac{m-m_H}{\epsilon})\hskip 20pt \text{for} \hskip 5pt m_H\in (m_0,m_n),
\end{equation}
\begin{equation}
\label{signal}
s_{\emptyset}(m)=0,
\end{equation}
where $c_{m_H}$ is a scaling constant, and $\phi$ is the normal probability density function with standard deviation $\epsilon$ which controls the spread of the signal function.

The use of finely binned data is common in the physics literature since the size of the data collected is often large. The likelihood of the binned data is given by,
\begin{equation}
\label{likelihood}
\pi(\mathbf{y}|\Lambda, m_H)= \prod_{i=1}^n \frac{\exp(-\Gamma_i)\Gamma_i^{y_i}}{y_i!},
\end{equation}
where
\begin{equation}
\label{bin_parameter}
\Gamma_i=\int_{m_{i-1}}^{m_i}[\Lambda(m)+s_{m_H}(m)]dm.
\end{equation}
The grid $\mathbf{m}=(m_0,m_1,\ldots,m_n)$ is the vector of bin boundaries over the search window. In practice, we treat $\int_{m_{i-1}}^{m_i}\Lambda(m)dm$ as observations of a log-Gaussian process to avoid integration of the log-Gaussian process at every evaluation of the likelihood in our computations. 

The posterior distribution of the model parameters $\boldsymbol{\theta}=(\eta,\sigma^2,\Lambda,m_H)$ given the data $\mathbf{y}$ can be written as
\begin{equation}
\label{post}
\pi(\boldsymbol{\theta}\mid\mathbf{y})=\frac{\pi(\boldsymbol{\theta})\pi(\mathbf{y}|\boldsymbol{\theta})}{\int \pi(\boldsymbol{\theta})\pi(\mathbf{y}|\boldsymbol{\theta})d\boldsymbol{\theta}},
\end{equation}
where $\pi(\boldsymbol{\theta})=\pi(\eta)\pi(\sigma^2)\pi(m_H)\pi(\Lambda)$ is the independent prior. The prior distribution, $\pi(m_H)$ is chosen as a mixture of a point mass at $m_H=\emptyset$ and a continuous distribution on $(m_0,m_n)$. In Section~\ref{calibration}, we argue that, after calibration, decision making under the proposed procedure is not sensitive to the choice of this prior. The hyperparameters $\eta$ and $\sigma^2$ are assigned inverse Gamma hyperpriors with shape and scale parameters equal to one. The hyperpriors prevent these parameters from moving to boundary values while allowing them to explore the plausible range of values.


Markov chain Monte Carlo (MCMC) sampling to infer the parameters of the above model is challenging with the mixture prior on $m$. Also because the likelihood is sensitive to small changes in the parameters and $\Lambda(m)$ is a function. To overcome computational difficulties we use an SMC algorithm \citep{Moral06}. The SMC samplers are a family of algorithms that take advantage of a sequence of distributions that bridge between a distribution that is straightforward to sample from (for example the prior) and the target distribution. Particles are filtered through the defined sequence using importance sampling and re-sampling steps to eventually obtain a sample from the target distribution. In a common version of SMC the sequence of filtering distributions is defined by tempering the likelihood, i.e., the role of likelihood is induced in the model in a sequential manner. In this case, SMC is especially useful in sampling $m$ from a mixture distribution.

Let the filtering sequence of distributions be denoted by,
$$\pi_0, \pi_1,\ldots,\pi_T.$$
Using a temperature schedule $\{\tau_t,t=0,\ldots,T\}$, the $t^{\text{th}}$ distribution in the sequence is defined as a power posterior,
$$\pi_t=\pi(\boldsymbol{\theta})[\pi(\mathbf{y}|\boldsymbol{\theta})]^{\tau_t},$$
where
$$0=\tau_0<\tau_1<\ldots<\tau_T=1.$$
The SMC sampler comprises iterative steps of weighting and sampling. While the particles are moved toward the target distribution through re-sampling with weights calculated according to the current temperature, they are also moved toward higher probability regions under each distribution in the sequence, through a sampling step, to prevent particle degeneracy. These steps are explained in Algorithm~\ref{alg1}. 

The form of the incremental weights $\tilde{w}_i$ depends on the choice of the transition kernel $K_t$ in SMC. In Algorithm 1 $K_t$ is chosen as an MCMC transition kernel that results in the simplified form of the  incremental weights. For more details about the weight calculation and sampling step in the SMC sampler see \cite{Moral06}. In Section~\ref{sec:data-analysis}, we explain our specific choices of the inputs of the algorithm where we apply the above Bayesian hierarchical model to the simulated Higgs data set.

\begin{algorithm}[t]
	\caption{Sequential Monte Carlo}\label{alg1}
	\begin{algorithmic}[1]
	\renewcommand{\algorithmicrequire}{\textbf{Input:}}
	\renewcommand{\algorithmicensure}{\textbf{Return:}}
	\Require A temperature schedule $\{\tau_t,t=0,\ldots,T\}$

           A MCMC transition kernel $K_t$\\
	
	Generate an initial sample $\boldsymbol{\theta}^{1:N}_0\sim \pi_0$;\\
	$W_1^{1:N}\gets \frac{1}{N}$;
	\For{$t:=1, \ldots, T-1$} 
	\begin{itemize}

	 \item $W_t^i\gets W_{t-1}^i\frac{\tilde{w}_t^i}{\sum \tilde{w}_t^i }$ where $\tilde{w}_t^i=P(\mathbf{y}|\boldsymbol{\theta}_t^{(i)})^{\tau_t-\tau_{t-1}}$, $i=1,\ldots,N$;
	
           \item Re-sample the particles $\boldsymbol{\theta}_t^{1:N}$ with importance weights $W_t^{1:N}$;
           
           \item $W_t^{1:N}\gets \frac{1}{N}$;
	
	\item Sample $\boldsymbol{\theta}_{t+1}^{1:N}\sim K_t$;
	
	\end{itemize}
	\EndFor

	\Ensure Particles $\boldsymbol{\theta}_T^{1:N}$.
	\end{algorithmic}
	\end{algorithm}

\section{Bayesian decision making}
In this section we consider the problem from a decision theoretic point of view. We define a linear loss function and derive the Bayes rule that can be used as an alternative to the current discovery/exclusion method for reporting one or more possible mass values for the Higgs particle. The Bayes procedure is calibrated to match specified frequency-theory-error-rates.

\subsection{Structure}
The required ingredients of a decision theory problem are a model with the corresponding parameter space, a decision space which is a set of possible actions to take, and a loss function \citep{Berger80}. 

The model was introduced in Section~\ref{model}. However, the procedure we now suggest could be used regardless of the specific details of the model. We define the decision space as the set of all possible subsets, $S\subset \cal{M}$ where $\mathcal{M}$ was defined in Section~\ref{model} as the union of the interval $(m_0,m_n)$ and the single point $\emptyset$ that represents the case that the Higgs particle does not exist. The interpretation of $S$ is that, $m\in S$ if, having observed the data, we wish to retain $m$ as a possible value of the true mass. For instance, if $\emptyset\in S$, the results suggest that it is possible that the Higgs particle does not exist (at least not with a mass in the search window). 

The next step is to define a loss function that reflects two goals. First we would like to include the correct parameter value in the decision set. Therefore we charge a penalty if the correct value is excluded from the decision set. Second, we would like to exclude from $S$ any incorrect parameter value. So we charge a penalty for including any parameter value that is not the true value. For the time being, suppose that the parameter space, ${\cal M}$, and the decision set, $S$, are discrete. Let $l(m_i)$ and $C(m_i)$ denote the loss values that respectively correspond to the case where $m_i$ is not the mass of the Higgs particle ($m_i\neq m_H$) but included in the decision set and the case where $m_i$ is the true mass of the Higgs particle ($m_i=m_H$) but excluded from the decision set. We refer to $l(m_i)$ and $C(m_i)$ as inclusion and exclusion losses respectively. Allowing $l$ and $C$ to depend on $m_i$ permits us later on to adjust these functions to give desired error rates. 

The following linear loss function accounts for all the possible decision scenarios with the corresponding losses,
\begin{equation}
\label{loss}
L_D(m_H,S)=\sum l(m_i) \mathbbm{1}(m_i\in S)\mathbbm{1}(m_i\neq m_H)+C(m_H)\mathbbm{1}(m_H\notin S)
\end{equation}
where the subscript $L_D$ shows the momentary discreteness assumption. However, we wish to treat the mass as a continuous variable over the interval of interest. Moreover, the case $m_H=\emptyset$ needs to be displayed explicitly in the formulas since it is treated differently than the rest of the parameter space in terms of error rates. Therefore, we begin by rewriting the loss function as
\begin{align}
\label{loss1}
L_D(m_H,S)= \Big[&\sum_{m_i\in S\cap (m_0,m_n)} l(m_i) \mathbbm{1}(m_i\neq m_H) 
\nonumber\\ &+l(\emptyset)\mathbbm{1}(\emptyset\in S)+C(m_H)\mathbbm{1}(m_H\notin S)\Big]\mathbbm{1}(m_H\in (m_0,m_n))\nonumber\\ 
+\Big[&\sum_{m_i\in S\cap (m_0,m_n)}l(m_i)\mathbbm{1}(m_i\neq m_H)+C(\emptyset)\mathbbm{1}(\emptyset\notin S)\Big]\mathbbm{1}(m_H=\emptyset)
\end{align}
To pass to the continuous case we now replace the sum over $S\cap (m_0,m_n)$ by an integral. (Conceptually, as we add more possible loss values we rescale the loss function to keep the total sum bounded. In the limit the function $l(m)$ on $(m_0,m_n)$ is a ``loss density"; it has units of ``loss'' per unit of mass.  The quantity $l(\emptyset)$ has units of ``loss'' as do the quantities $C(\emptyset)$ and $C(m_H)$.) The result is the loss function
\begin{align}
\label{loss2}
L(m_H,S)=\Big[&\int_{S\cap(m_0,m_n)}l(m)\mathbbm{1}(m\neq m_H)dm\nonumber\\ &+l(\emptyset)\mathbbm{1}(\emptyset\in S)+C(m_H)\mathbbm{1}(m_H\notin S)\Big]\mathbbm{1}(m_H\in (m_0,m_n))\nonumber\\ 
+\Big[&\int_{S\cap(m_0,m_n)}l(m)dm+C(\emptyset)\mathbbm{1}(\emptyset\notin S)\Big]\mathbbm{1}(m_H=\emptyset)
\end{align}
The indicator $\mathbbm{1}(m\neq m_H)$ can be dropped from the first integral without changing its value to give the  simplified form
\begin{align}
\label{loss}
L(m_H,S)=&\int_{S\cap(m_0,m_n)}l(m)dm+\left[l(\emptyset)\mathbbm{1}(\emptyset\in S)+C(m_H)\mathbbm{1}(m_H\notin S)\right]\mathbbm{1}(m_H\in (m_0,m_n)) \nonumber\\ 
&+C(\emptyset)\mathbbm{1}(\emptyset\notin S)\mathbbm{1}(m_H=\emptyset),
\end{align}
where the term $\int_{S\cap (m_0,m_n)}l(m)dm$ is the loss due to including incorrect mass values in $S$.


By averaging the loss function (\ref{loss}) with respect to the marginal posterior $\pi(m\mid\mathbf{y})$ the posterior expected loss or the Bayes risk is obtained as follows.
\begin{align}
r_{\pi(m|\mathbf{y})}(S)&=E_{\pi(m|\mathbf{y})}[L(m,S)]\nonumber\\
&=\int_{S\cap(m_0,m_n)} l(m)dm+\int_{S^c\cap(m_0,m_n)}C(m)\pi(m\mid\mathbf{y})dm\nonumber\\
&\hskip 10pt +C(\emptyset)\mathbbm{1}_{S^c}(\emptyset)\pi(\emptyset\mid\mathbf{y})+l(\emptyset)\mathbbm{1}_S(\emptyset)\left(1-\pi(\emptyset\mid\mathbf{y})\right)
\end{align}
The Bayes rule is obtained by minimizing the Bayes risk with respect to $S$.  

\newtheorem{thm}{Theorem}
\begin{thm}
\label{thm1}
The Bayes rule, i.e., the decision rule that minimizes $r_{\pi(m|\mathbf{y})}(S)$, is given by
\begin{equation}
S=\begin{cases}\{m\in (m_0,m_n): \frac{l(m)}{C(m)}<\pi(m\mid \mathbf{y}) \} & \frac{l(\emptyset)}{C(\emptyset)}\ge\frac{\pi(\emptyset\mid \mathbf{y})}{1-\pi(\emptyset\mid \mathbf{y})}
\\
\{m\in (m_0,m_n): \frac{l(m)}{C(m)}<\pi(m\mid \mathbf{y}) \} \cup\{\emptyset\}& \frac{l(\emptyset)}{C(\emptyset)}<\frac{\pi(\emptyset\mid \mathbf{y})}{1-\pi(\emptyset\mid \mathbf{y})} .\end{cases}
\end{equation}
\end{thm}
A proof is provided in Appendix~\ref{proofofthm}.

\subsection{Calibration and error rate estimation} \label{calibration}
As mentioned before, the proposed procedure can be calibrated to give desired frequency theory
properties such as error rates. The loss ratios $
l(m)/C(m)$ can be adjusted to satisfy the type I error rates required in particle physics applications. The same effect could be achieved in principle by keeping the loss ratio fixed and adjusting the prior. However, fixing the prior makes the computations below more straightforward. The global Type I error rate and false exclusion rates are controlled respectively as follows.
\begin{align}\label{eqn:calib_1}
P(\emptyset\notin S\mid m=\emptyset)&=P\left(\frac{\pi(\emptyset\mid\mathbf{y})}{1-\pi(\emptyset\mid\mathbf{y})}<\frac{l(\emptyset)}{C(\emptyset)}\mid m=\emptyset\right)\nonumber\\
&=\alpha_1,
\end{align}
\begin{align}\label{eqn:calib_2}
P(m\notin S\mid m)&=P\left(\pi(m\mid\mathbf{y})<\frac{l(m)}{C(m)}\mid m\right)\nonumber\\
&=\alpha_2.
\end{align}
where $P(A)$ is the probability of event $A$.  

The Associate Editor has observed that these calibrations are not strictly frequentist in nature because of the way we are handling the background $\Lambda$.  A strict frequentist procedure would have 
\begin{align}\label{eqn:calib_1_alt}
P(\emptyset\notin S|m=\emptyset,\Lambda)&=\alpha_1,
\\
\label{eqn:calib_2_alt}
P(m\notin S|m,\Lambda)&=\alpha_2.
\end{align}
Such calibration is likely impossible if we insist on exact calibration in small samples.  Traditional statistical procedures such as likelihood ratio tests are calibrated by letting the critical value for the test statistic depend on the nuisance parameters. In other words, equation~(\ref{eqn:calib_1_alt}) is required to hold only at the estimated value $\hat\Lambda$ of $\Lambda$; approximate calibration is then achieved by parametric bootstrapping.  The procedure we describe below is parallel but averages over those $\Lambda$ which remain credible after seeing the data, i.e., according to the posterior distribution of $\Lambda$; details are given in the next two subsections.  Our procedure and parametric bootstrapping are both properly calibrated in large samples.
  
Note that calibration to produce desired frequency error rates offsets the effect of the prior in decision making. Since the posterior odds can be written as a product of the prior odds and likelihood ratio, the  loss ratio we obtain and use
in decision making, responds to the choice of the prior. 

Solving equations~(\ref{eqn:calib_1}) and (\ref{eqn:calib_2}) for the loss ratios $l(\emptyset)/C(\emptyset)$ and $l(m)/C(m)$ requires obtaining the $\alpha_1 100\%$ and $\alpha_2100\%$ quantiles of the null distributions of the posterior odds 
$\pi(\emptyset|\mathbf{y})/(1-\pi(\emptyset|\mathbf{y}))$, 
and the posterior probability density, $\pi(m|\mathbf{y})$. 

Unfortunately, under most realistic models the distribution of the posterior functionals cannot be obtained in closed form. \cite{Johnson13} developed the uniformly most powerful Bayesian test (UMPBT) for one-parameter exponential family based on the same idea, i.e., maximizing the probability that the Bayes factor is smaller than a certain threshold under the null model. They briefly visit the Higgs problem and report the size of a Bayes factor equivalent to the local significance level of $\alpha_0=3\times 10^{-7}$. However, to be able to obtain the UMPBT, a normal model is used and the cross section is treated as an unknown parameter as it is in the existing analysis of the Higgs data. 

The results in \cite{Johnson13} cannot be used under our model. Therefore, we need  to estimate percentiles of the distribution of the posterior using Monte Carlo. However, this requires intense computation since for each generated data set at each iteration of the Monte Carlo we need to run the SMC algorithm to estimate the posterior. This Monte Carlo within Monte Carlo scheme is computationally costly on its own, while satisfying the small significance level in the physics application requires a large number of iterations to estimate precise tail quantiles adding to the computational intensity.

To address calibration with affordable computation, we combine importance sampling and approximation techniques: we replace the SMC algorithm with a Laplace approximation in each Monte Carlo algorithm and use importance sampling to reduce the number of iterations required to obtain tail probability estimates for the Bayesian statistic's distribution for a fixed level of precision.

\subsubsection{Approximation}
\label{approx}
In the following, we explain the Laplace approximation to the marginal posterior distribution of the mass of the Higgs particle. This approximation is used as a fast alternative to sampling the posterior distribution to speed up the calibration Monte Carlo. The hyperparameters $\eta$ and $\sigma^2$ are held fixed at their maximum a posteriori estimates in the calibration. Consider reparametrizing the model in terms of $\Psi=\log \Lambda$. The approximation method, inspired by \cite{Rue09}, is based on a Gaussian approximation to the conditional distribution, $\tilde{\pi}(\Psi\mid m,\mathbf{y})$, i.e.,
\begin{equation}
\label{eqn:approx}
\pi(m\mid\mathbf{y})=\left. \frac{\pi(m,\Psi,\mathbf{y})}{\tilde{\pi}(\Psi\mid m,\mathbf{y})}\right |_{\Psi=\Psi^*}.
\end{equation}
The Gaussian approximation, $\tilde{\pi}(\Psi\mid\mathbf{y},m)$, is obtained by numerically approximating the mode and curvature of $\pi(\Psi\mid\mathbf{y},m)$;
\begin{equation}
\pi(\Psi\mid\mathbf{y},m) \propto \exp\{-\frac{1}{2}(\Psi-\boldsymbol{\mu})^T\Sigma^{-1}(\Psi-\boldsymbol{\mu})+\log \pi(\mathbf{y}|\Psi , m)\}.
\end{equation}
Consider the Taylor expansion of the $n$ components of the log likelihood around the initial values $\Psi_0$,
\begin{align}
\log \pi(\mathbf{y}|\Psi , m)&=\sum_{i=1}^{n}g_i(\Psi_i)\nonumber \\ 
&\approx \sum_{i=1}^n [g_i(\Psi_{0i})+g'_i(\Psi_{0i})(\Psi_i-\Psi_{0i})+\frac{g''_i(\Psi_{0i})}{2}(\Psi_i-\Psi_{0i})^2] \nonumber\\ 
&= \sum_{i=1}^{n} [a_i(\Psi_{0i})+b_i(\Psi_{0i})\Psi_{i}-\frac{1}{2}c_i(\Psi_{0i})\Psi_{i}^2].
\end{align}
where, 
\begin{align}
&g_i(\Psi)=\log(\pi(y_i\mid \Psi, m)), \label{g}\\ &a_i(\Psi_{0i})=g_i(\Psi_{0i})-g'_i(\Psi_{0i})\Psi_{0i}+\frac{g''_i(\Psi_{0i})}{2}\Psi_{0i}^2,\\ &b_i(\Psi_{0i})=g'_i(\Psi_{0i})-\Psi_{0i}g''_i(\Psi_{0i}),\\
&c_i(\Psi_{0i})=-\frac{g''_i(\Psi_{0i})}{2}\label{c}.
\end{align}
The above expressions are given explicitly in Appendix~\ref{ApB}. Therefore, we have
\begin{align}
\tilde{\pi}(\Psi\mid\mathbf{y},m)&\propto \exp\{-\frac{1}{2}(\Psi-\boldsymbol{\mu})^T\Sigma^{-1}(\Psi-\boldsymbol{\mu}) \nonumber\\
&\hskip 38pt +\sum_{i=1}^{n} [a_i(\Psi_{0i})+b_i(\Psi_{0i})\Psi_{i}-\frac{1}{2}c_i(\Psi_{0i})\Psi_{i}^2]\} \nonumber\\
&\propto  \exp\{ -\frac{1}{2}\Psi^T(\Sigma^{-1}+\text{diag}(\mathbf{c}_0))\Psi+(\Sigma^{-1}\boldsymbol{\mu}+\mathbf{b}_0)^T\Psi\}.
\end{align}
where $\mathbf{b}_0=(b_1(\Psi_{01}),\ldots,b_n(\Psi_{0n}))^T$ and $\mathbf{c}_0=(c_1(\Psi_{01}),\ldots,c_n(\Psi_{0n}))^T$. The mean (mode) of the approximate Gaussian distribution, $\tilde{\pi}(\Psi\mid\mathbf{y},m)$, is obtained by repeatedly solving $(\Sigma^{-1}+\text{diag}(\mathbf{c}_t))\Psi_{t+1}=(\Sigma^{-1}\boldsymbol{\mu}+\mathbf{b}_t)$ for $\Psi_{t+1}$ until convergence, where $\mathbf{b}_t$ and $\mathbf{c}_t$ are updated matrices at iteration $t$. The approximate covariance matrix is $\Sigma^{-1}+\text{diag}(\mathbf{c})$, where $\mathbf{c}$ is the matrix  $\mathbf{c}_t$ after convergence. Therefore, the approximate marginal distribution can be obtained up to a normalizing constant as follows,
\begin{align}
\label{approx-post}
\tilde{\pi}(m\mid\mathbf{y}) &\propto \pi(m)\int \tilde{\pi}(\Psi\mid m,\mathbf{y}) d\Psi\nonumber\\
&\propto\pi(m)|\Sigma^{-1}+\text{diag}(\mathbf{c})|^{-\frac{1}{2}}.
\end{align}

\subsubsection{Importance sampling for estimating error rates}
\label{IS}
As mentioned before, to evaluate the error rates associated with the Bayesian testing procedure, tail probabilities of the posterior functionals need to be estimated. Accurate Monte Carlo estimates for probabilities of rare events are only obtained with large Monte Carlo samples (in the order of $10^7$ and larger in this application). In this section we introduce an importance Monte Carlo algorithm that is used to obtain tail probability estimates with lower variances.

We focus on the global type I error rate of the Bayesian testing procedure, i.e.,
\begin{equation}
\alpha_1=P_{H_0}\left(\frac{\pi(\emptyset \mid \mathbf{y})}{1-\pi(\emptyset\mid \mathbf{y})}<\frac{l(\emptyset)}{C(\emptyset)}\right)=P(\pi(\emptyset\mid \mathbf{y})<q_\emptyset)
\end{equation}
where $q_\emptyset=l(\emptyset)/(C(\emptyset)+l(\emptyset))$. While in calibrating the Bayes procedure the goal is to estimate $q_\emptyset$ to satisfy a determined $\alpha_1$, suppose, for the time being,  that $\alpha_1$ is to be estimated for a given $q_\emptyset$.

To estimate $\alpha_1$ using basic Monte Carlo,  data, $\mathbf{y}_i$, is generated in each iteration under the null hypothesis, $H_0$, and $\pi(\emptyset\mid \mathbf{y}_i)$ is obtained. The Monte Carlo estimate of $\alpha_1$ based on a (large) sample of $N$  posterior values is given by,
\begin{equation}
\hat{\alpha}_1=\frac{1}{N}\sum_{i=1}^N \mathbbm{1}(\pi(\emptyset)\mid \mathbf{y}_i)\in (0,q_\emptyset))
\end{equation}
However, under the null hypothesis, the event that $\pi(\emptyset \mid \mathbf{y}_i)$ falls bellow $q_0$ is rare and an unaffordably large $N$ is required to obtain a non-zero estimate for $\alpha_1$. 

Importance sampling is a popular method for simulating rare events \citep{Rubino09}. The idea is to generate samples under an importance distribution under which the event of interest is likely to occur and weight the samples according to the original distribution of interest. To use importance Monte Carlo, here, we seek an importance distribution under which small values of $\pi(\emptyset\mid\mathbf{y})$ are more likely to occur. 

Let us remind ourselves of the model under the null and alternative hypotheses,
\begin{align}
&H_0:\text{The Higgs particle does not exist, i.e., }m_H=\emptyset,\\
&H_A: \text{The Higgs particle exists with a mass } m_H\in(m_0,m_n),
\end{align}
Clearly we expect the event $\pi(\emptyset\mid\mathbf{y})<q_\emptyset$ to occur with high probability under the alternative. Therefore we can use the model under $H_A$ as the importance distribution. The importance weights are then given by,
\begin{align}
\label{importance-weight}
W_i&=\frac{\pi(\mathbf{y}\mid H_0)}{\pi(\mathbf{y}\mid H_A)}\notag\\
&=\frac{\pi(\mathbf{y}\mid m=\emptyset)}{\int_{m_0}^{m_n}\pi(\mathbf{y}\mid m)dm}\notag\\
&=\frac{\pi(\emptyset\mid \mathbf{y})\pi(\mathbf{y})/\pi(\emptyset)}{\int_{m_0}^{m_n}[\pi(m\mid \mathbf{y})\pi(\mathbf{y})/\pi_A(m)]dm}\notag\\
&=\frac{\pi(\emptyset\mid \mathbf{y})}{\pi(\emptyset)\int_{m_0}^{m_n}[\pi(m\mid \mathbf{y})/\pi_A(m)]dm},
\end{align}
where $\pi(\emptyset)$ and $\pi_A(m)$ are the priors over the mass under $H_0$ and $H_A$, respectively. The importance Monte Carlo estimate of $\alpha_1$ based on a sample generated under the alternative model is given by,
\begin{equation}
\label{p-est}
\tilde{\alpha}_1=\frac{1}{N}\sum_{i=1}^N\mathbbm{1}_{(0,q_\emptyset)}(\pi(\emptyset\mid \mathbf{y}_i)))W_i.
\end{equation}
For calibration, however, (\ref{p-est}) is solved for $q_\emptyset$ with a given significance level, $\alpha_1$. Algorithm~\ref{calibration-alg} outlines the calibration steps. As mentioned earlier, the marginal posterior of the mass is obtained by integrating the background over its posterior distribution, i.e., the procedure is calibrated for the most likely realizations of the background function. This is done by repeatedly sampling from the posterior sample of the background generated by Algorithm~\ref{alg1} in step 1-b of Algorithm~\ref{calibration-alg}.

\begin{algorithm}[t]
\caption{Importance Monte Carlo calibration algorithm}\label{calibration-alg}
\begin{algorithmic}[1]
\renewcommand{\algorithmicrequire}{\textbf{Input:}}
\renewcommand{\algorithmicensure}{\textbf{Return:}}
\Require Pre-determined significance level, $\alpha_1$.
 
\For{$i:=0,1, \ldots, N$} 
\renewcommand{\labelitemi}{}

\begin{itemize}
\item[a-] Generate $m_i\sim \pi_A(m)$;

\item[b-] Generate $\boldsymbol{\Lambda}_i\sim \pi(\boldsymbol{\Lambda}\mid \mathbf{y}_{obs})$ (sample a realization from the posterior sample generated by Algorithm~\ref{alg1});

\item[c-] Generate data, $\mathbf{y}_i\sim \pi(\mathbf{y}\mid \boldsymbol{\Lambda}_i,m_i)$;

\item[d-] Obtain $\tilde{\pi}(m\mid \mathbf{y}_i)$ using~(\ref{approx-post});

\item[e-] Obtain $W_i$ using~(\ref{importance-weight}).
\end{itemize}
\EndFor

\item Solve~(\ref{p-est}) to obtain $q_0$.

\Ensure Discovery threshold $q_0$.
\end{algorithmic}
\end{algorithm}


\section{The Higgs simulated data analysis}\label{sec:data-analysis}
\subsection{Data}
Confidentiality rules do not allow access to the real data for non-members of the Higgs research groups. In this section we apply our model and procedure to simulated data provided to us by Matthew Kenzie of the CMS group and described in \cite{CMS14} and \cite{CMS-PAS-HIG-13-001}.  The simulation procedure is very complex because it must model not only the predicted behavior of the Higgs boson but also the behavior of the extremely complex CMS detector. Analysis of such simulated data was an essential step in developing the analytic techniques to be used for the real experiment.
The simulated data available to us represent the diphoton decay mode invariant mass spectrum (in the range $100 < m_{\gamma\gamma} < 180$ GeV) at centre of mass energy $\sqrt{s}=8$TeV. For each of the decay modes there are different Higgs signatures referred to as analysis categories. For the diphoton decay mode there are nine analysis categories. We had access to a list of the invariant mass of each data event together with the corresponding analysis category. 

There are several ``production mechanisms" through which a Higgs particle can be generated; five such mechanisms were considered for our data. Each such production mechanism leads to a specific predicted signal function. While the production mechanism is not identified in the data the signal function is propagated through the analysis separately for each of the SM Higgs production mechanisms at the LHC and each analysis category.  The shape of the signal function in each analysis category, for each of the production modes and at three hypothesized Higgs masses (120, 125, and 130 Gev) has been provided to us in form of a histogram; the entry in a single bin of such a histogram is  the expected number of Higgs events produced by a specific production mechanism in a specific analysis category in the mass range for that bin if the Higgs has the particular hypothesized mass. A handful of the production mechanism, analysis category combinations produce so few expected outcomes that we were not provided the corresponding signal histograms; in the end we have histograms for 41  of the 45 combinations.

\begin{figure}[t]
  \centering
    \includegraphics[width=.8\textwidth]{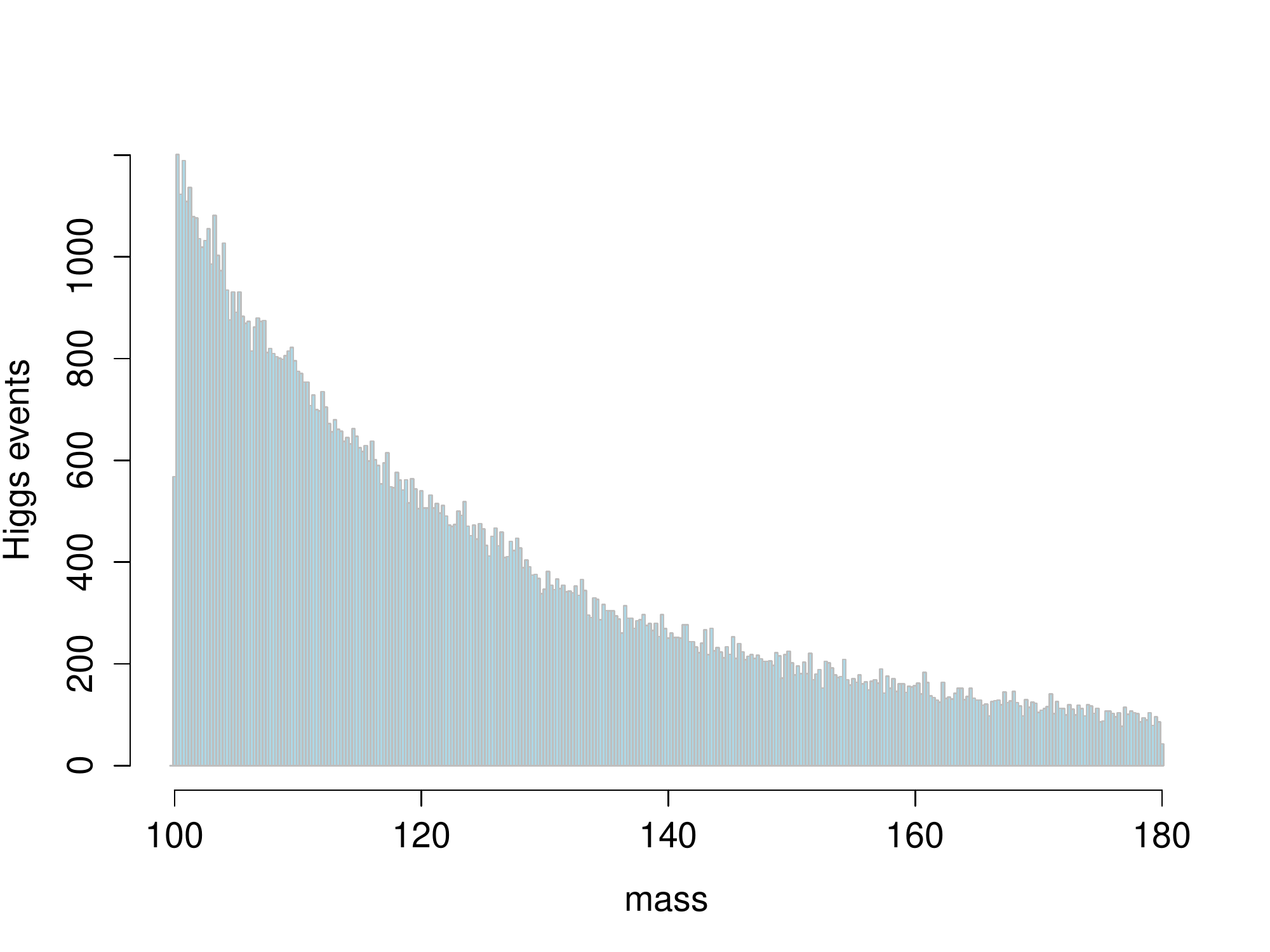}
  \caption{Simulated data representing the invariant masses of events}
\label{Higgs-data}
\end{figure}

We fit the signal function,~(\ref{signal}), to these histograms and estimate the signal strength, $c_m$, and signal width, $\epsilon$. We then extrapolate from the three signal masses we were given to obtain the signal function corresponding to other masses. We pool the data for all nine analysis categories and bin the data according to the signal histograms with 322 bins. Figure~\ref{Higgs-data} shows the histogram of the data. The reason for pooling the analysis categories is that some of these categories have very few data points and that the computational burden of a full analysis exceeded our capabilities. We also use the sum of the signal functions over the production modes and analysis categories as a single signal function for each mass. 

Note that, in principle, our method can handle the complications produced by having several combinations of production mechanism/decay mode/analysis category. However, simulated data is provided to us for only one decay mode and in practice, low event counts in some of the analysis categories can be a source of problems in making inference separately in these categories.

\subsection{Inference}\label{sec:data_inference}
Following the model introduced in Section~\ref{model}, the background is modelled as a log-Gaussian process with known prior mean. The prior mean for the background function is chosen to be a fourth-order Bernstein polynomial function that is a typical parametric form in the current practice,
\begin{equation}
\label{polynomial-background}
\xi(z)=\sum_{i=0}^{4}b_ig_i(z),
\end{equation}
where $z\in (0,1)$ is an affine transformation of mass ($m\in(m_0,m_n)$), into the unit interval. This transformation helps specification of the correlation parameter, $\eta$, in the spatial covariance function, (\ref{covfnct}). The basis functions, $g_i(z)$, are given by,
\begin{equation}
g_i(z)=\binom{4}{i}z^i(1-z)^{4-i},
\end{equation}
and $b_i$, the polynomial coefficients are selected by fitting the polynomial to the data. In practice, the known mean function can be obtained from Monte Carlo background-only data as is typically done to determine a parametric model for the background \citep{Phys4}. Here, in the absence of any other source of information to elicit the prior mean, we use a polynomial fit to the data. This choice resembles empirical Bayes where the maximum likelihood of the hyperparameters are used to determine the prior. A fixed mean function allows us to use assumptions that are typically made about the background function while letting the data decide if these assumptions are legitimate: the log-Gaussian process with diffuse hyperpriors on the covariance parameters $\eta$ and $\sigma^2$ allows deviations from the mean function if the data contains information about such deviations. 

\begin{figure}[t]
  \centering
    \includegraphics[width=.8\textwidth]{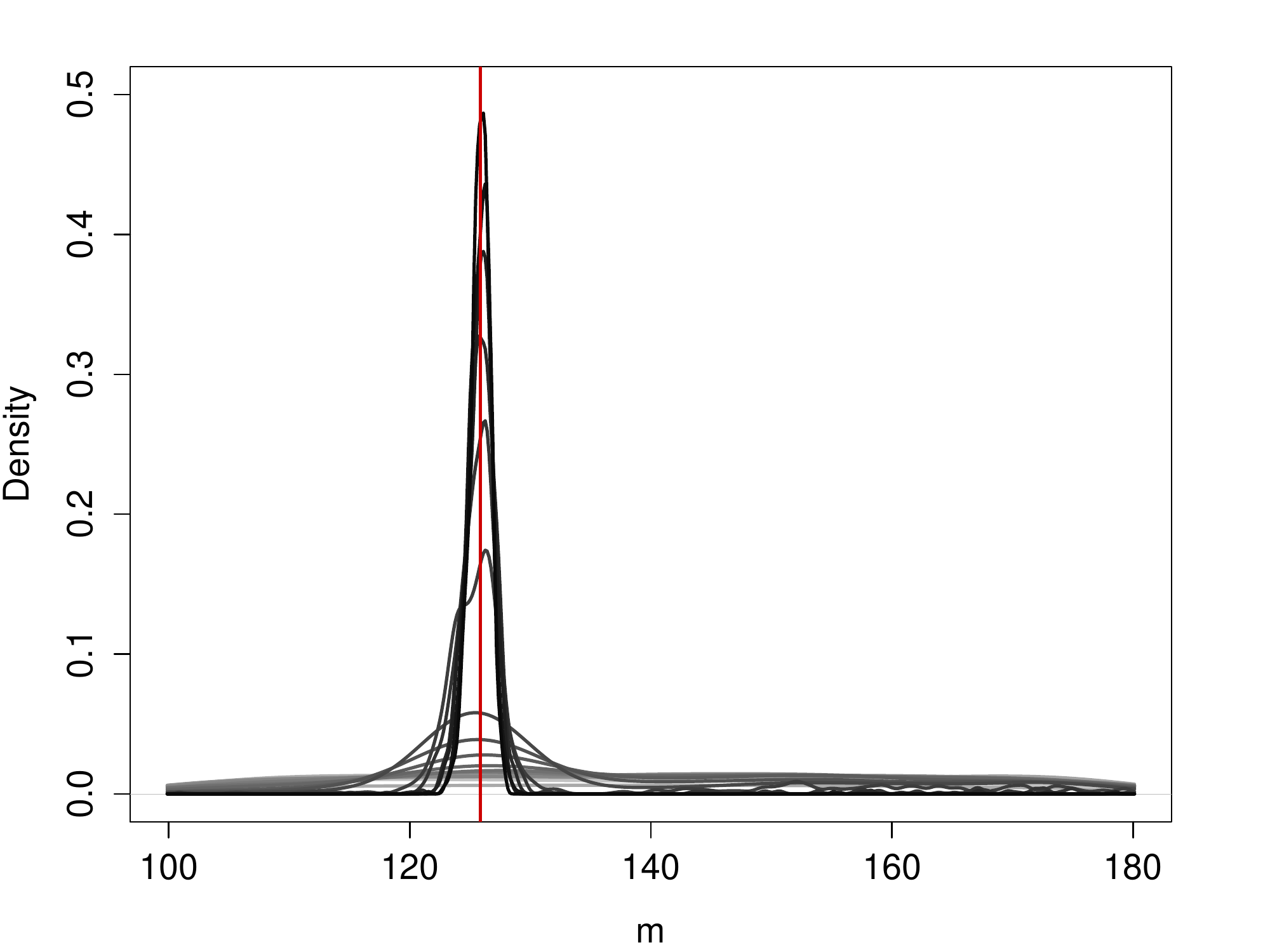}
  \caption{The results of the analysis of the simulated data representing the posterior density of the mass evolving through the sequential sampler as the likelihood is induced into the power posterior sequentially. The estimated mass of the Higgs particle ($\hat{m}_H\approx 126$) is specified by the vertical red line.}
\label{m_seq}
\end{figure}

\begin{figure}[t]
  \centering
    \includegraphics[width=.8\textwidth]{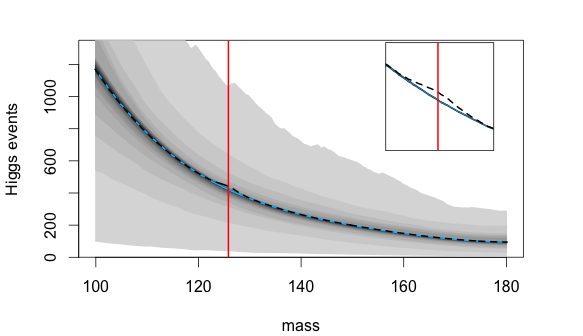}
  \caption{The results of the analysis of the simulated data representing the target posterior for the background and the posterior background mean plus the signal function centred at the maximum a posteriori mass. The estimated mass of the Higgs particle ($m_H\approx 126$) is specified by the vertical red line.}
\label{back_seq}
\end{figure}

The prior distribution defined for the parameter $m$ is a mixture distribution given by,
\begin{equation}
\pi(m_H)=\begin{cases} 0.5 & m_H=\emptyset\\ \frac{0.5}{m_n-m_0} & m_H\in (m_0,m_n).\end{cases}
\end{equation}
This is a ``non-informative" prior given the original formulation of the problem as testing the hypothesis $H_0:$ ``no Higgs", against $H_1:$ ``Higgs with mass $m_H\in(m_0,m_n)$". This prior assigns half of the probability mass to each model. We also tried a ``uniform" prior that assigns equal probability to all values of $m$ including  $m=\emptyset$ that resulted in posterior samples consistent with the one presented in the paper (Figures~\ref{m_seq} and \ref{back_seq}). 

The sampling step in Algorithm~\ref{alg1} is performed by generating the background and mass from proposal distributions followed by an accept/reject step. The proposal distribution for the background at time $t+1$ is a log-Gaussian distribution with mean equal to the background at time $t$ and covariance matrix following the prior covariance structure but scaled according to the posterior variances at time $t$. The proposal distribution for $m$ is determined adaptively by estimating the marginal distribution of $m$ given $\mathbf{y}$ at time $t$ as 
\begin{equation}
\hat{\pi}_t(m\mid \mathbf{y})=\begin{cases} \frac{\sum_{i=1}^N\mathbf{1}(m_t^i=\emptyset)}{N} & m=\emptyset\\
 \frac{\sum_{i=1}^N\mathbf{1}(m_t^i\neq\emptyset)}{N}\zeta(m) & m\neq \emptyset.\end{cases}
\end{equation} 
where $\zeta(m)$ is a kernel density estimate of the distribution of $m\neq \emptyset$ at time $t$. The temperature schedule, $\{\tau_t\}$, is chosen to be a grid of size 20 on $[0,1]$.

The results, in terms of the posterior samples of mass and background curves over the 20 steps of the sequential algorithm, are presented in Figures~\ref{m_seq} and~\ref{back_seq} respectively. Figure~\ref{m_seq} shows kernel density estimates of the posterior distribution of mass, $m_H$, along the sequence of densities. The colour of the curves get darker as the sample is trimmed and moved towards the target posterior. The red vertical line shows the maximum a posteriori mass value  ($\hat{m}_H\approx 126$). In Figure~\ref{back_seq} the gray bands show the 95\% credible intervals for the background function; the wide bright bands refer to the early steps of the algorithm where curves are sampled from the diffuse prior with covariance parameters that result in non-smooth and highly variable curves while as the credible bands get darker they become smoother and more focused about the target posterior mean. 

%
%



\subsection{Unknown cross section}\label{sec:unknown_mu}
In our model introduced in Section~\ref{model} and the analysis of the simulated data above, we use the cross section of the Higgs particle predicted by the SM. In this section we discuss a version of our Bayesian hierarchical model in which the uncertainty about the cross section is incorporated. This model is defined by replacing the signal function $s_{m_H}$ in (\ref{bin_parameter}) by $\mu s_{m_H}$ where $\mu$ is a unitless parameter similar to the one used in the discovery stage of the current procedure. However, we note that our inference results for the parameter $\mu$ are not comparable to those in the literature (e.g. this report\footnote{\href{https://twiki.cern.ch/twiki/bin/view/CMSPublic/Hig12020TWiki}{https://twiki.cern.ch/twiki/bin/view/CMSPublic/Hig12020TWiki}}). The reason is that our signal function does not match the one used in the current analysis of the Higgs data. Moreover, we multiply the overall signal function, i.e.,  the sum of the signal functions over the production modes and analysis categories, by parameter $\mu$ rather than fitting the cross section separately for each production mode. We emphasize once more that this could be done in our framework in principle but since the simulated data contains too few data points in some of the production mode/analysis categories we implement the model only for the data aggregated over all the  combination categories.

In the Bayesian framework, the cross section uncertainty can be incorporated by adding the corresponding parameter to the model together with a prior distribution. However, with free parameters in both the background and signal functions inference can be more challenging since the posterior surface is likely to be multimodal: by letting the signal strength vary the likelihood may take the same value for two very different scenarios: $\mu \approx 0$ with a background function with a short length scale (small correlation) that fits any small bump in the data; and $\mu > 0$ with a smooth background. To avoid such situations we use an informative prior for the cross section. We use a log-normal prior distribution for $\mu$ with parameters 0 and 0.05 taking most of its probability mass between 0.85 and 1.2. 

\begin{figure}[t]
	\centering
	\includegraphics[width=.8\textwidth]{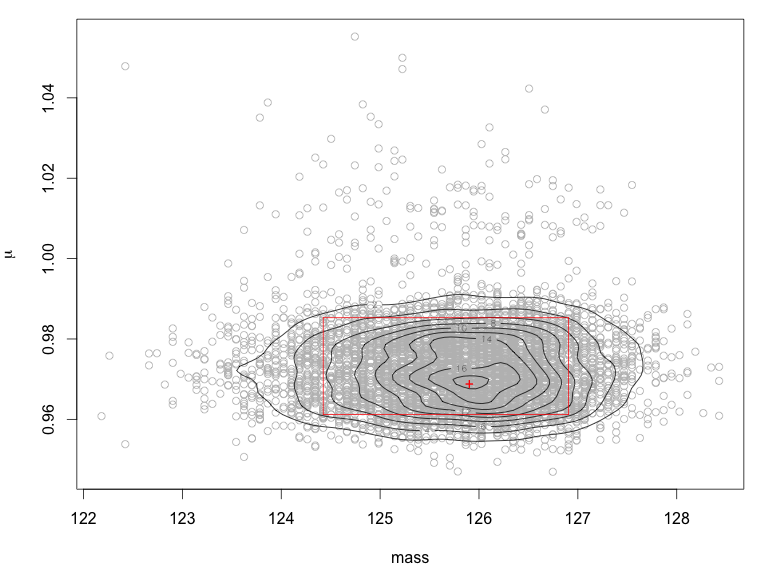}
	\caption{The joint posterior samples of the mass and the cross section parameter. The cross shows the joint posterior mode estimated by kernel density estimation. The contours are the kernel density estimates of the joint posterior. The rectangle is the box that contains the top 68\% of the posterior sample.}
	\label{fig:mu_mass}
\end{figure}

The results of fitting this model to the simulated data are presented in Figure~\ref{fig:mu_mass} in form of samples and estimated contours of the joint posterior distribution of the mass and the cross section parameter. The estimate of the joint posterior mode is specified with the cross and the box contains the top 68\% of the joint posterior. As mentioned earlier the value of the estimate for the cross section parameter is meaningless and pointless to compare with estimates given in previous analysis since, due to lack of access to real data and signal information, we have made a number of arbitrary choices here. Instead, the purpose of this section is  to demonstrate the capability of the Bayesian approach to adapt to the uncertain cross section scenario. 

\subsection{Calibrated discovery threshold and the decision set}
We run Algorithm~\ref{calibration-alg} for the model introduced in Section~\ref{model} to obtain the discovery threshold.  The lower $(3\times10^{-7})$100\% quantile of the null distribution of the posterior probability, $\pi(\emptyset\mid \mathbf{y})$ is $q_0=0.0066$. The discovery decision based on the simulated data, $\mathbf{y}_{sim}$ is made by comparing the estimated null posterior probability to the obtained threshold. The SMC-based point estimate of $\pi(\emptyset\mid \mathbf{y}_{sim})$ is zero, i.e., no samples with $m=\emptyset$ remain in the final posterior sample of size 5,000. A  conservative 90\%  upper limit on this estimate would be about 0.0063 which is smaller than the discovery threshold. Therefore, we conclude that the simulated data contains adequate evidence of the existence of the Higgs particle.

%

\begin{figure*}[t]
\centering
\begin{subfigure}[b]{0.47\textwidth}
                \centering
                \includegraphics[width=\textwidth]{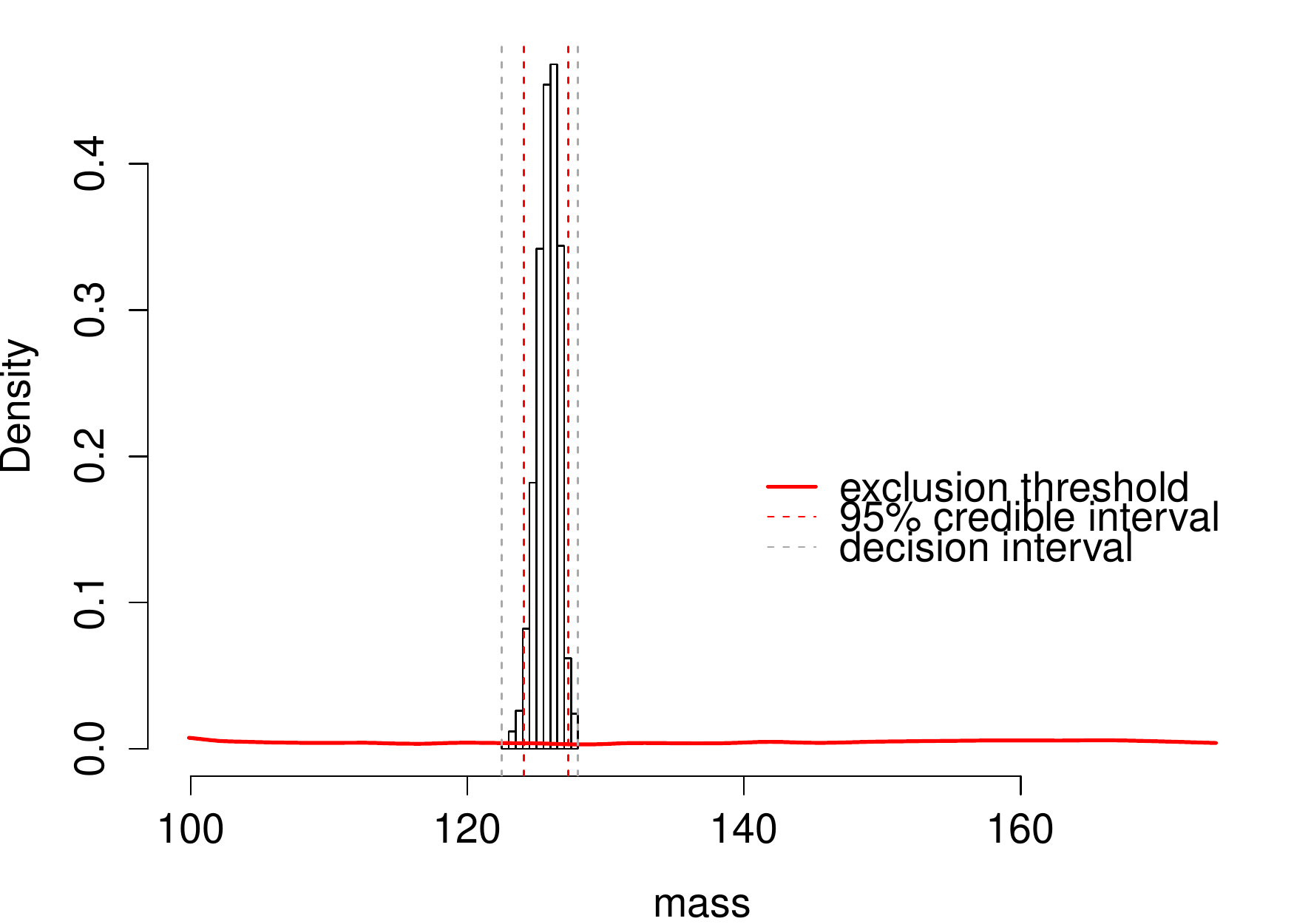}
               \caption{}
                \label{exclusion_threshold_post}
        \end{subfigure}
 \begin{subfigure}[b]{0.47\textwidth}
                \centering
                \includegraphics[width=\textwidth]{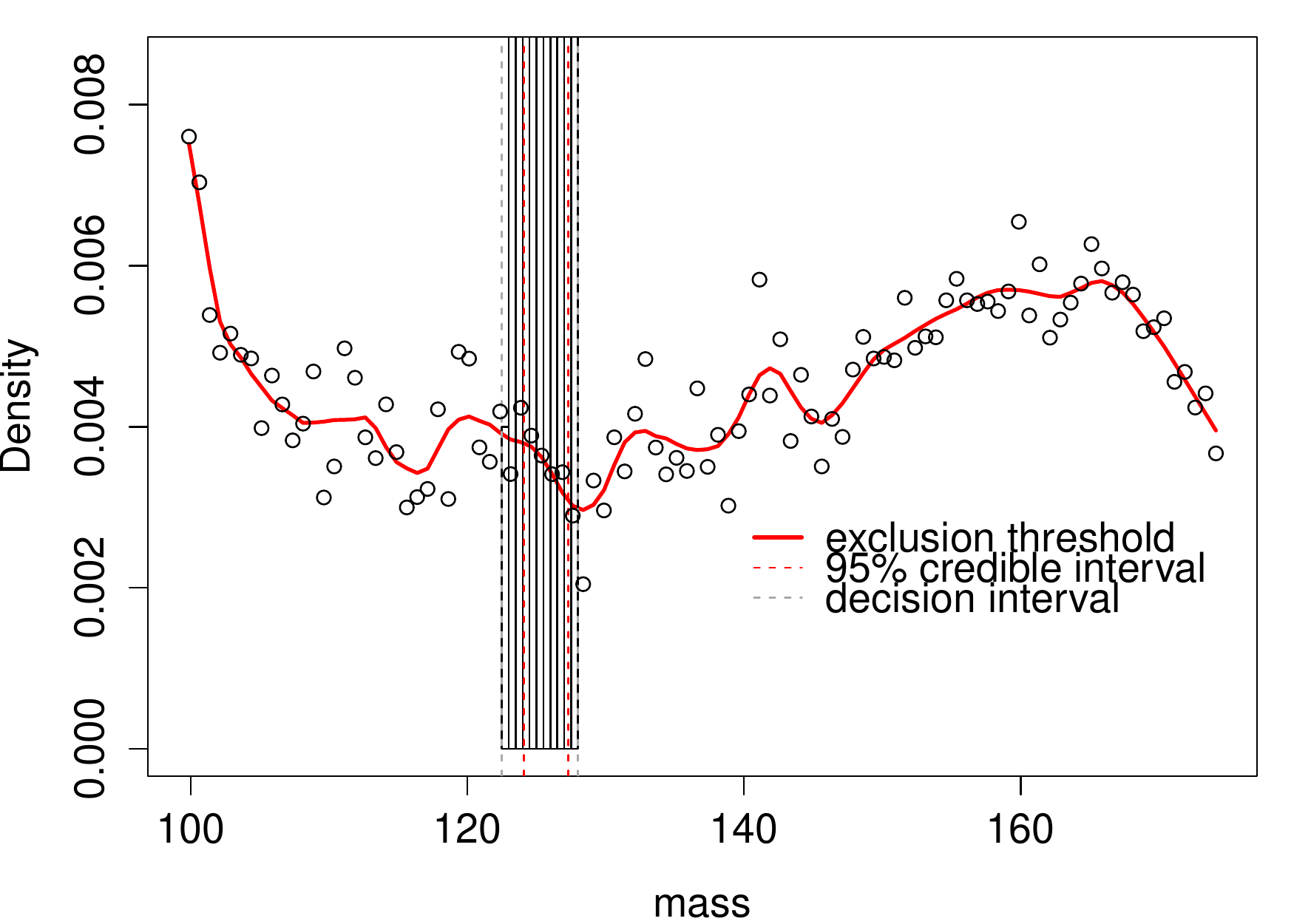}
               \caption{}
                \label{exclu_thresh}
        \end{subfigure}
\caption{Exclusion threshold plotted over the histogram of the posterior sample of mass together with the decision and Bayesian credible intervals. In (b) the y axis is limited to small densities for better visualization, the points are the decision thresholds obtained for the coarse grid over the mass.}\label{ex_thresh}
\end{figure*}

Having concluded that $\emptyset\notin S$, to obtain the final decision set we only need to obtain the exclusion thresholds, $q(m)$, for $m\in(100,180)$, such that,
\begin{equation}
P(\pi(m\mid\mathbf{y})<q(m)\mid m)=\alpha_2,
\end{equation}
where $\alpha_2=0.05$ is a common choice. Obtaining the exclusion thresholds is analogous to that of discovery threshold given in Section~\ref{calibration}. Since the exclusion controlled error rates are not as small as the discovery thresholds even basic Monte Carlo can provide accurate estimates. However, the approximation in (\ref{eqn:approx}) is pointwise and the computational burden increases with the size of the mass grid. Therefore, the total computational time for the exclusion step is $O(N^2\times M)$ for a mass grid of size $N$ and $M$ Monte Carlo iterations. To reduce the computational cost and given that the exclusion step is not as sensitive as the discovery step we use a coarse discretization of the mass spectrum and use a kernel smooth of the exclusion thresholds for this coarse grid. Figure~\ref{ex_thresh} shows the exclusion threshold together with the histogram of the posterior sample of mass in the search window. The boundaries of the decision set are determined by the mass values where the estimated posterior density is higher than the exclusion threshold. These boundaries are specified in Figure~\ref{ex_thresh} by the vertical dashed grey lines. In addition to the decision interval we provide the 95\% Bayesian credible interval, denoted by $S_B$ that satisfies
\begin{equation}
\int_{S_B}\pi(m\mid\mathbf{y})=0.95.
\end{equation}
An estimate of $S_B$ is obtained as an interval whose lower and upper limits are given by the 0.025 and 0.975 sample quantiles of the posterior over mass: $\hat{S}_B=(124.26,127.46)$ shown by the vertical dashed red lines. The decision set obtained by the exclusion threshold cuts is $\hat{S}=(122.5,128)$ which is more conservative than the Bayesian credible set.

\section{Conclusion and Discussion}
We have proposed a Bayesian procedure that can be used as an alternative to the detection/exclusion method used in the search for the Higgs particle. In a decision theoretic framework, we define a linear loss function that summarizes the possible outcomes of search for a new particle with the associated losses. The Bayes rule is obtained by minimizing the expected loss and is the basis of decision making. 

We introduce a  Bayesian hierarchical model to make Bayesian inference about the mass of the Higgs particle (the parameter of interest) as well as the background function. A sequential Monte Carlo algorithm is used to obtain samples from the posterior distributions of the model parameters. The model is fit to data produced by computer models that simulate the behaviours of the detectors. The analysis results are consistent with the analysis of the real Higgs data reported in physics literature - the posterior distribution of the mass is peaked about the reported mass of the Higgs particle \citep{phys1,phys2}. The 95\% credible Bayesian interval is reported as the set of credible values of the Higgs boson based on this analysis. 

Our Bayes procedure can be calibrated to give required frequency-theory-error-rates. A calibration algorithm is proposed in which the posterior distributions are obtained by a fast Laplace approximation instead of SMC thereby making the calibration step computationally feasible. An importance Monte Carlo for simulation of rare events is proposed that enables us to calibrate the Bayes procedure to meet the small type I error rate typically used in particle physics. We calibrate our procedure according to the error rate requirements in particle physics. A decision set is reported that excludes the null hypothesis (the hypothesis of ``no Higgs") and contains a range of masses that remain plausible according to discovery and exclusion error rates of $3\times 10^{-7}$ and $0.05$ respectively. 


An interesting observation is the similarity of the nature of our procedure to that of \cite{PhysRevD.57.3873} for constructing confidence intervals that was brought to our attention by a referee. \cite{PhysRevD.57.3873} discuss the coverage issues of confidence intervals that are built based on the two step procedure without proper conditioning and propose a method for constructing confidence intervals based on ranking the ratio of the likelihood under the null and alternative models. Our method is  similar to this approach in that the posterior odds are used for constructing the decision set.

We now revisit the criticisms/questions that were mentioned in Section~\ref{sec:intro} about the current practice and explain  what makes our proposed methodology different and preferable. In regard to the treatment of the cross section parameter, we introduce our model with the theoretical value of the cross section and argue that if the information is provided by the theory it should be used consistently in the search procedure as oppose to the current practice that makes use of this information only at the exclusion step. On the other hand if there is uncertainty associated with the cross section given by the theory, this uncertainty must be incorporated into the model. We address the global versus local error rates issue by defining a decision making procedure that can be calibrated to meet global frequency-theory-error-rates while under the current practice the local type I error rates are controlled and the global error rate is yet to be quantified. 

Another issue with the current practice is the independent local hypothesis testing over the mass spectrum in which it is unclear what the decision would be if more than one local hypothesis test are significant in the discovery step. Under the proposed procedure, however, discovery is only associated with the exclusion of $m=\emptyset$ in the decision set. Then any mass interval that remains in the decision set is considered as possible mass of the Higgs particle. It is worth mentioning that detecting well-separated peaks in the posterior of the mass is possible under our procedure.

\section*{Acknowledgments}
We are grateful to Louis Lyons and Matthew William Kenzie of the CMS group who provided to us the simulated data as well as insight on the problem throughout the course of this work.

\newpage
\begin{appendices}

\linespread{1}
\section{Proof of Theorem~\ref{thm1}}
\label{proofofthm}

\begin{proof}
Consider the problem of deciding whether or not mass value $m^*\in{\cal M}$ should be included in the decision set. Following the Bayes rule, $m^*$ is included if the change in risk associated with inclusion of $m^*$ is negative. To avoid measure theoretic complications we restrict the decision space to finite unions of open intervals and $\{\emptyset\}$. First, consider the case that $m^*\in(m_0,m_n)$.  Since the prior distribution of $m$ on the interval $(m_0,m_n)$ is absolutely continuous, the addition of any point like $m^*$ (as well as any other zero measure set) to the decision set leaves the risk unchanged. Let $S$ be the decision set before including $m^*$. We consider the change in the risk as a result of adding the interval $(m^*-\delta,m^*+\delta)$ for very small $\delta$ to $S$. Let $S_0=S\cup (m^*-\delta,m^*+\delta)$. We include $m^*$ in the final decision set if and only if,
\begin{equation}
r(S_0)-r(S)<0,
\end{equation}
where
\begin{align}
r(S_0)-r(S)&=\int_{S\cap (m_0,m_n)} l(m)dm+\int_{m^*-\delta}^{m^*+\delta}l(m)dm+\int_{S^c\cap (m_0,m_n)}C(m)\pi(m\mid\mathbf{y})dm\\
& -\int_{m^*-\delta}^{m^*+\delta}C(m)\pi(m\mid \mathbf{y})dm-\int_{S\cap (m_0,m_n)} l(m)dm-\int_{S^c\cap (m_0,m_n)}C(m)\pi(m\mid\mathbf{y})dm\\
&=\left\{l(m^*)-C(m^*)\pi(m^*\mid \mathbf{y})\right\}2\delta+o(\delta).
\end{align}
The above expression is negative for all sufficiently small $\delta>0$ if and only if
\begin{equation}
l(m^*)-C(m^*)\pi(m^*\mid \mathbf{y})<0,
\end{equation}
or
\begin{equation}
\frac{l(m^*)}{C(m^*)}<\pi(m^*\mid \mathbf{y}).
\end{equation}
For the case that $m^*=\emptyset$, let $S_0=S\cup\{\emptyset\}$,
\begin{align}
r(S_0)-r(S)&=\int_{S\cap (m_0,m_n)} l(m)dm+l(\emptyset)\left(1-\pi(\emptyset\mid \mathbf{y})\right)+\int_{S^c\cap (m_0,m_n)}C(m)\pi(m\mid\mathbf{y})dm\\
&\hskip 10pt -C(\emptyset)\pi(\emptyset\mid\mathbf{y})-\int_{S\cap (m_0,m_n)}l(m)dm-\int_{S^c\cap (m_0,m_n)}C(m)\pi(m\mid\mathbf{y})dm\\
&=l(\emptyset)\left(1-\pi(\emptyset\mid \mathbf{y})\right)-C(\emptyset)\pi(\emptyset\mid\mathbf{y}).
\end{align}
Therefore, $0$ is added to $S$ if
\begin{align}
l(\emptyset)\left(1-\pi(\emptyset\mid \mathbf{y})\right)-C(\emptyset)\pi(\emptyset\mid\mathbf{y})&<0\\
\frac{l(\emptyset)}{C(\emptyset)}<\frac{\pi(\emptyset\mid \mathbf{y})}{1-\pi(\emptyset\mid \mathbf{y})}.
\end{align}
\end{proof}

\section{Likelihood Expansion Terms - explicit expressions for (\ref{g}-\ref{c})}\label{ApB}
\begin{equation}
g_i(\Psi)=-\exp(\Psi)-s_i+y_i\log(\exp(\Psi)+s_i)-log(y_i!),
\end{equation}
where,
\begin{equation}
s_i=\int_{m_{i-1}}^{m_i}s_{m_H}(m)dm.
\end{equation}
\begin{align}
a_i(\Psi)=&g_i(\Psi)-g_i'(\Psi)\Psi+\frac{g''_i(\Psi)}{2}\Psi^2 \nonumber \\ \nonumber
=& -\exp(\Psi)-s_i+y_i\log(\exp(\Psi)+s_i)-log(y_i!)\\ \nonumber
&+\left(\exp(\Psi)-y_i\frac{\exp(\Psi)}{\exp(\Psi)+s_i}\right)\Psi\\ 
&+\left(-\exp(\Psi)+y_i\frac{\exp(\Psi)s_i}{(\exp(\Psi)+s_i)^2}\right)\Psi^2
\end{align}
\begin{align}
b_i(\Psi)=&g_i'(\Psi)\Psi+\Psi g''_i(\Psi) \nonumber \\ 
=& -\exp(\Psi)(1-\Psi)+y_i\frac{\exp(\Psi)}{\exp(\Psi)+s_i}(1-\frac{\Psi s_i}{\exp(\Psi)+s_i})
\end{align}
\begin{align}
c_i(\Psi)=&-\frac{g''_i(\Psi)}{2} \nonumber \\ 
=& \frac{1}{2}\exp(\Psi)(1-y_i\frac{\exp(\Psi)s_i}{(\exp(\Psi)+s_i)^2})
\end{align}
\end{appendices}

\newpage
\bibliographystyle{apalike}
\bibliography{refs}

\end{document}